# Computer-Generated Holographic Gratings in Soft Matter

*Gianluigi Zito[1,2], Antigone Marino[1,3], Bruno Piccirillo[1,2], Volodymyr Tkachenko[3], Enrico Santamato[1,2], Giancarlo Abbate[1,3,4]*

[1]University Federico II of Naples, Physics Dpt.
zito@na.infn.it
[2]CNISM, Consorzio Nazionale Interuniversitario per le Scienze fisiche della Materia c/o Physics Dpt. University of Naples "Federico II"
[3]CNR-INFM - Coherentia c/o Physics Dpt. University of Naples "Federico II"
*via Cinthia, 80126 Naples - Italy*
[4] CNR-INFM Licryl - Liquid Crystal Laboratory c/o Physics Dpt. University of Calabria
*87036 Rende (CS), Italy*

*Standard multiple-beam holography has been largely used to produce gratings in polymer-liquid crystal composites, like POLICRYPS, H-PDLC gratings and POLIPHEM* [1]. *In this work we present a different approach to liquid crystal-polymeric grating production, based on the Computer-Generated Holography (CGH). The great advantage of using CGH is that interferometer-based schemes are no longer necessary, avoiding problems related to long term stability of the interference pattern and multi-beam complex optical setup. Moreover, the CGH technique allows a wider choice of pattern designs. In this preliminary work, we obtained promising results, as for instance the patterning of a square-wave refractive index modulation of a LC-polymeric composite, a pattern which is not achievable with standard two-beam holography.*

**INTRODUCTION**

In recent years there has been an increasing interest in the optical properties of complex dielectric materials. On one side, a large research effort is dedicated to periodic structures, as photonic crystals [2]. On the other side, many groups are studying the interesting properties of complex dielectric structures known as quasi-crystals [3]. Furthermore, adding optical gain to both classes of materials has allowed observing laser action [4], opening the way to new efficient laser devices with a number of attractive applications.

Soft materials like polymers, liquid crystals and composites, as substrate to be patterned in complex geometries, exhibit very attracting optical, mechanical, chemical and thermal properties and flexible processing techniques. Combined with holographic fabrication techniques, they can be patterned quite easily into several unique structures. The record of the holographic interference pattern usually takes the form of the spatial modulation of the absorption constant or the refractive index of the medium, or both. Photo-polymeric materials yield modulation of the refractive index [5]. Standard two-beam holography produces only sinusoidal modulations in soft matter gratings [6], whereas patterning of complex structures requires a dual-beam multiple exposure technique [7], or $N$ laser beams to create a 2-D $N$-fold quasi-periodic irradiance profile, as it has been realized with holographic lithography [8]. Both techniques require complex optical setups with drawbacks related to the long term stability and vibrational control of the interference pattern.

In past years a new kind of electrically switchable holographic grating has been recorded in liquid crystal-polymer composite materials, usually known as H-PDLC (holographic-polymer-dispersed liquid crystals) [1, 6]. H-PDLCs are typically produced by curing, under a laser interference pattern, an isotropic photosensitive liquid mixture containing, basically, a pre-polymeric material, a liquid crystal and a photo-initiator. On the other hand, Computer-Generated Holograms (CGHs), in conjunction with a Spatial Light Modulator (SLM), permit to create arbitrary 2-D and 3-D configurations of single-beam irradiance distribution in coincidence of the Fourier plane of the reconstructing lens [9]. Moreover, commercially available SLMs offer nanometer-scale spatial resolution and real-time reconfigurability without any mechanical motion or realignment [10].

In this work, we used a single-beam computer-generated holography to functionalize liquid crystal-polymer composites. We analyzed the potentialities of the CGH method realizing an experimental setup for optical writing of several structures. With our apparatus, we were able to obtain, for instance, a square-wave modulation of the refractive index of a liquid crystal-polymer composite. This patterning is not achievable with standard two-beam holography.

## EXPERIMENTAL SETUP AND RESULTS

The scheme of the experimental setup that we employed for curing the photosensitive mixtures is shown in Fig. 1. The Gaussian input laser beam (at $\lambda = 532nm$) is expanded and spatially filtered before it impinges on the spatial light modulator, so to utilize entirely its modulating surface and ensuring the best efficiency. The CGHs are addressed as intensity images with 256 levels of grey to the SLM display. Each level of grey of the hologram, addressed to the pixel at $\mathbf{r}_{ij}$, corresponds to a certain voltage that reorients the liquid crystal molecules, resulting in a different, discrete, value of the optical phase $\Phi_{in}(\mathbf{r}_{ij})$. The digitalized holographic profile $\Phi_{in}(\mathbf{r}_{ij})$ was computed with an iterative Adaptive-Additive algorithm (AA) [11, 12]. Each step of the iteration requires a two-dimensional (direct/inverse) Fast Fourier Transform routine [13, 14]. The AA algorithm solves the problem of the CGH-calculation exploring, numerically, the space of degenerate phase profiles which can encode the desired irradiance profile $I_f(\boldsymbol{\rho})$ at the back focal plane of the microscope objective, where the sample is placed. Due to the Fourier transform properties, any displacement of the hologram plane, produced, for instance, by environmental vibrations, does not affect the irradiance profile. Our LC-spatial light modulator is the HoloEye Photonics LC-R 3000. The system is based on a high resolution WUXGA LCoS display of $1200 \times 1920$ pixels. The CGH can be addressed at a maximum frame rate of $120Hz$. The pixel size is $l_{in}=9.5\mu m$, whereas the fill factor of the display is 92% on an array of $19.01 \times 11.40 mm^2$, with a phase modulation ranging between 0 and $2\pi$. We took into account possible experimental deviations from the ideal reproduction of the optimal hologram by adding uncorrelated Gaussian noise profiles to the calculated phase profile [12].

The phase-only hologram in the *H* input plane is transferred to the input pupil of the microscope objective, in the *H\** conjugate plane, by means of the relay lenses (Fig. 1). The large focal length (150-200$mm$) of the lenses minimize spherical aberrations and coma. They permit also a fine adjustment of the hologram size in the *H\** plane and hence of the grating spacing in the *FT* focal plane, with a sensitivity of about $100nm$. Several holograms $\Phi_{in}(\mathbf{r}_{ij})$ were computed to realize different irradiance profiles $I_f(\boldsymbol{\rho})$, from a simple sinusoidal modulation or standard square 1-D modulation to more complex irradiance distributions. Figure 1 shows, on the bottom, two examples of our holograms, obtained by adaptive-additive algorithm. The CGH in Fig. 1-(a) provides the irradiance profile represented in Fig. 1-(c) (photo acquired with a CCD camera), which shows a one-dimensional periodic square modulation with different sizes of the dark and bright fringes. The CGH in Fig. 1-(b) provides the 2-D quasi-periodic Penrose irradiance pattern, represented in Fig. 1-(d), which presents an 8-fold rotational symmetry.

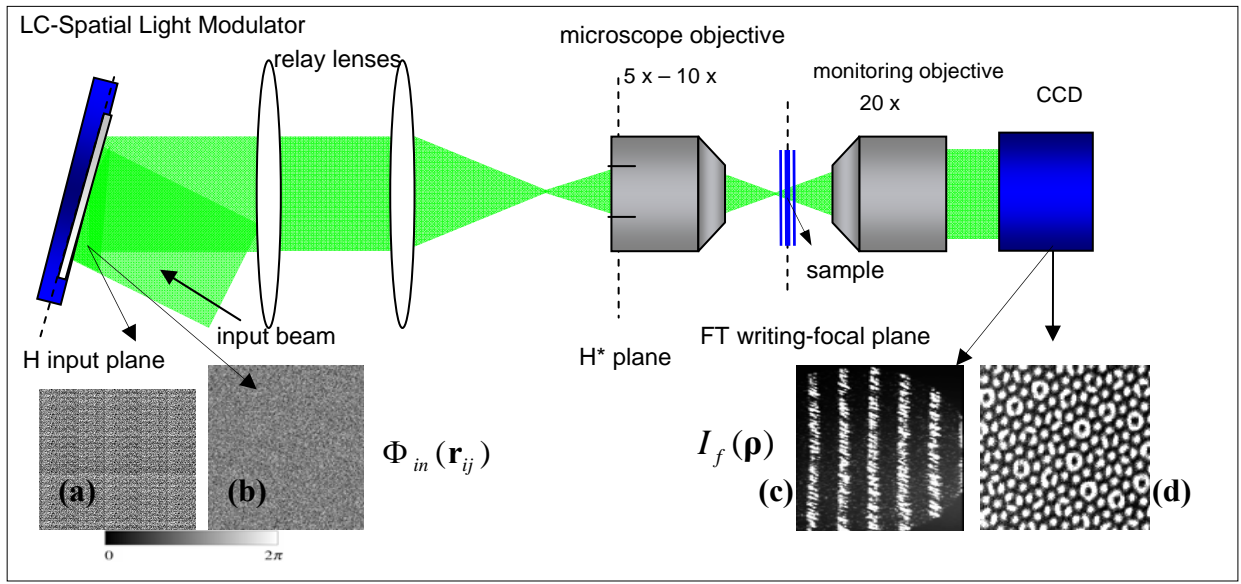

**Figure 1.** The laser beam (a continuous wave Nd:YVO$_4$ frequency-doubled Coherent Verdi), operating at a wavelength $\lambda = 532 nm$, after being expanded and spatially filtered, impinges on a reflective SLM reproducing, in the *H* input plane, the CGH $\Phi_{in}(\mathbf{r}_{ij})$. Then the beam is projected by the relay lenses onto the conjugated plane *H\**, which lies onto the input pupil of the microscope objective. Hence, the writing irradiance pattern $I_f(\mathbf{\rho})$ is reconstructed in the *FT* writing-focal plane. A second microscope objective permits the real-time monitoring of the writing pattern with a CCD camera. Figures (a) and (b) show the phase-only holograms computed to design, respectively, the irradiance profile (c) and (d): a 1-D square pattern with different sizes of the dark and bright fringes (c), and a Penrose 8-fold irradiance profile (d).

The overall structure can be changed in real-time resizing the hologram via computer. By varying the scale factor of the phase profile $\Phi_{in}(\mathbf{r}_{ij})$ with a magnification factor $\alpha$, in fact, the length-scale in the focal plane of the microscope objective changes with an inverse magnification factor $\alpha^{-1}$. The effect of resizing the hologram on a 1-D square irradiance profile is shown in Fig. 2-(a-b). The grating pitch of the light pattern was changed from $30 \mu m$ (a) to $15 \mu m$ (b) by enlarging the original relative CGH of a scale factor of two.

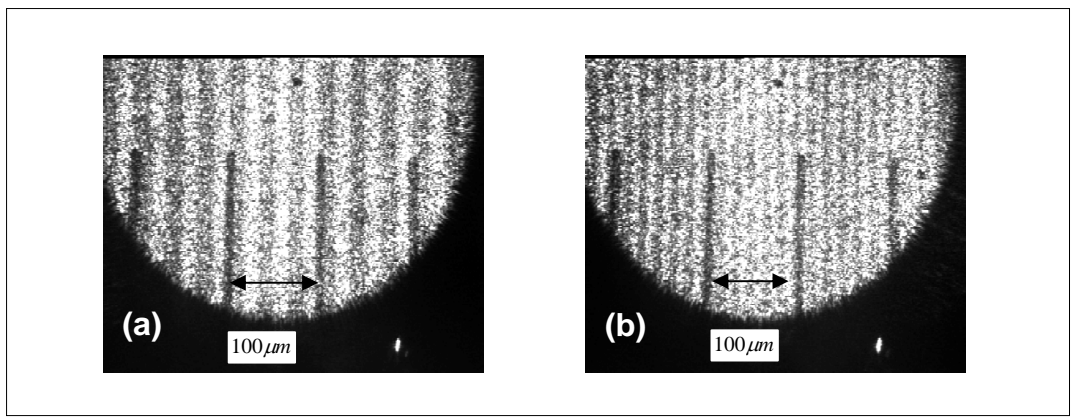

**Figure 2**. The effect of resizing the holograms: the grating pitch of a square-wave irradiance profile $I_f(\mathbf{\rho})$ was changed from $30\mu m$ (a) to $15\mu m$ (b) by enlarging the original CGH of a scale factor of two.

Our gratings were obtained by curing, at room temperature, a pre-polymer/LC photosensitive mixture in the focal plane of a microscope objective, where the desired writing irradiance profile is reconstructed. We used a starting solution of the monomer dipentaerythrol-hydroxyl-penta-acrylate DPHPA (60.0% w/w), the cross-linking stabilizer monomer N-vinylpyrrolidinone (9.2% w/w), the liquid crystal BLO38 by Merck (30.0% w/w), and a mixture of the photoinitiator Rose Bengal (0.3% w/w) and the coinitiator N-phenylglycine (0.5% w/w) [15]. Our first attempts concerned patterns with 1-D square and sinusoidal modulations. Two examples of gratings written into LC-polymeric mixtures are shown in Fig. 3-(a-b). The square-wave polymeric grating shown in Fig. 3-(a) corresponds to the irradiance distribution shown in Fig. 2-(a).

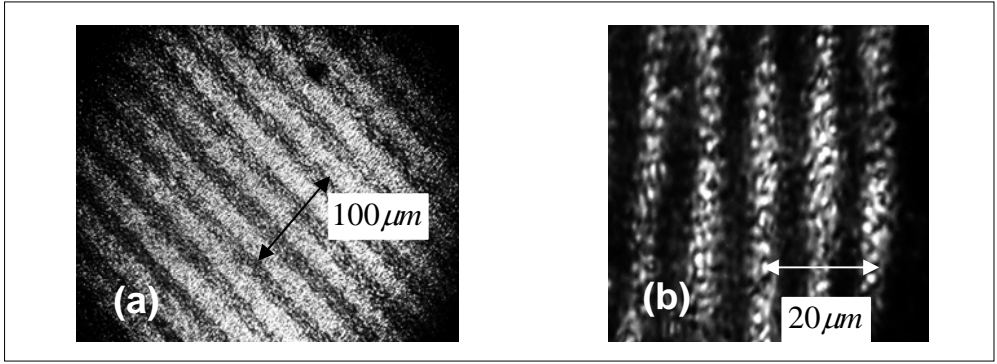

**Figure 3**. The images (a-b) show the result of the square-wave modulations of the LC content, and hence of the average refractive index, obtained by optical curing of our pre-polymer/LC mixtures with our CGH-patterns $I_f(\mathbf{\rho})$ (the photos are taken under a polarizing microscope).

A similar grating with a smaller period $\Lambda = 12\mu m$ is shown in Fig. 3-(b). Limited by the optical setup, the cured area had, typically, a linear dimension of $\approx 1mm$, whereas the intensity of the writing beam was of the order of $2 \div 10 mW/cm^2$ at the sample position. The exposure times ranged from about $30s$ to $300s$.

According to Whittaker-Shannon sampling theorem [13], the pixel size $l_f$ in the *FT* plane is related to the pixel size $l_{in}$ in the input plane in accordance with the relation $l_f = \frac{\lambda f}{N l_{in}}$, where *f* is the focal length of the reconstructing lens and $N$ represents the total number of pixels on the hologram sides. The quite large values of the grating spacing obtained in our preliminary writing patterns $I_f(\mathbf{\rho})$ are not a limitation of the CGH technique. In fact, the modulation in the irradiance profile, realized with several pixel sizes $l_f$ (typically $1 \div 2\mu m$), was produced with a number of pixels per fringe of $10 \div 20$, whereas the microscope objectives employed as the reconstructing lens had a $5\times$-$10\times$ magnification. Using a larger magnification and irradiance designs with larger resolution could offer a larger spatial resolution up to nanometer-scale of the writing patterns [12], although with the drawback of smaller cured areas.

At the present, the limitations in writing our irradiance patterns on the photosensitive mixtures are due to the low contrast of the intensity distributions $I_f(\mathbf{\rho})$, which were typically 4:1 and only in a few cases rise to 25:1. This is related, mainly, to the low diffraction efficiency of our CGHs $\Phi_{in}(\mathbf{r}_{ij})$, which was typically less than 40%, in our first experimental tests, and thus to the large amount of non-diffracted light, which was mixed-up to the diffraction pattern, lowering its contrast. Moreover, the limited depth of field of our microscope objective (about $60 \mu m$) made critical the task of correctly positioning the writing plane and the sample cell (the cell thicknesses ranged from a few microns to tens of microns). Furthermore, the presence of speckle (aliasing) [14] in the irradiance profile also contributed to decrease the overall quality of the writing patterns. Overcoming these experimental difficulties is, at the moment, the more remarkable problem we are dealing with. Possible solutions, presently under consideration, could be improving the CGH-calculation algorithm (to reduce speckle), and mainly separating the diffracted light from the non-diffracted beam (to improve the contrast).

**CONCLUSIONS AND PERSPECTIVES**

We demonstrated, in this work, the real possibility of writing dielectric modulation in soft matter composites using computer-generated holograms. We highlighted the main advantages coming from this single-beam holographic technique. The choice of the pattern design is almost without limitation. Besides the very attractive possibilities of such unlimited choice, this writing technique avoids the necessity of a multi-beam optical setup and consequent mechanical control and optics realignment. In fact, the CGH technique is quite insensitive to environmental vibrations. We computed several high resolution holograms ($1200 \times 1920$ pixels) with an FFT-based iterative adaptive-additive algorithm. The efficiency of the generated irradiance profiles in inducing a dielectric modulation was verified on several LC/polymeric samples. Our preliminary attempts succeeded in producing square-wave phase gratings, which are not achievable with standard two-beam holography. We are now dealing with the production and optical characterization of these and more complex structures.

These results are very preliminary and further improvements are under way. Our present activity is focussed on the optimization of the computer-generated holograms, i.e. reducing the aliasing (speckle) with zero-padding or tiling techniques [14], increasing the diffraction efficiency, improving the contrast and the overall quality of the writing pattern. Further planning concerns the realization of devices with complex 1-D and 2-D quasi-periodic patterns.